\documentclass[aps,prl,showpacs,twocolumn,a4paper]{revtex4}
\usepackage{graphicx}
\usepackage{multirow}

\begin{document}

\title{Redistributing Chern numbers of Landau subbands: tight-binding electrons under staggered-modulated magnetic fields}
\author{Yi-Fei Wang$^1$ and Chang-De Gong$^{2,1}$}
\affiliation{$^1$National Laboratory of Solid State
Microstructures and Department of Physics,
Nanjing University, Nanjing 210093, China\\
$^2$Department of Physics, The Chinese University of Hong Kong,
Hong Kong, China}
\date{\today}

\begin{abstract}
We investigate the magneto-transport properties of electrons on a
square lattice under a magnetic field with the alternate flux
strength $\phi\pm\Delta\phi$ in neighboring plaquettes. A new
peculiar behavior of the Hall conductance has been found and is
robust against weak disorder: if $\phi={p\over{2N}}\times{2\pi}$
($p$ and $2N$ are coprime integers) is fixed, the Chern numbers of
Landau subbands will be redistributed between neighboring pairs
and hence the total quantized Hall conductance exhibits a direct
transition by $\pm{N}e^2/h$ at critical fillings when $\Delta\phi$
is increased from $0$ up to a critical value $\Delta\phi_c$. This
effect can be an experimental probe of the staggered-flux phase.
\end{abstract}

\pacs{73.43.Cd, 75.47.-m, 73.20.Fz, 71.10.Hf} \maketitle

{\it Introduction.---}The integer quantum Hall effect (IQHE) is
observed in two-dimensional (2D) electron systems, in the presence
of a strong perpendicular uniform magnetic field~\cite{Hall}. The
phenomenology of IQHE is well explained by Landau quantization and
disorder-induced localization. The IQHE electronic states exhibit
interesting topological properties~\cite{Thouless,Arovos}. In
particular, each state can be labelled by a topologically
invariant integer called the Chern number, which is the Hall
conductance of this state, in units of $e^2/h$. A state with a
nonzero Chern number carries a Hall current and is necessarily
extended. In solving the problem regarding how an IQHE state
transforms into the insulating state, many authors carried out
numerical studies of the tight-binding model in the presence of a
uniform magnetic field and a random
potential~\cite{Liu,Yang1,Sheng,Yang2}. An interesting behavior
has been reported by Sheng and Weng that disorder will induce
direct higher IQHE plateau to insulator transition~\cite{Sheng}.
These theoretical results are in good agreement with several
magneto-transport experiments performed in Si
metal-oxide-semiconductor field-effect
transistors~\cite{Kravchenko} and a 2D hole system confined in a
Ge/SiGe quantum well~\cite{Song}.

Recently, an intriguing orbital-current-carrying state in the
square lattice, the staggered-flux phase (SFP) which is also known
as orbital antiferromagnet or $d$-density wave (DDW)
state~\cite{Chakravarty}, regains new attentions since unusual
experimental findings in two systems: a pseudogap phenomena (See
the review by Timusk {\it et al.}\cite{Timusk}) in the underdoped
region of high-$T_c$ cuprates~\cite{Chakravarty}; a hidden order
in the heavy-fermion compound URu$_{2}$Si$_{2}$~\cite{Chandra}.
However, a direct experimental observation of the SFP order is
difficult. On the other hand, adopting a superconducting wire
network decorated with an array of ferromagnets, an artificial
system of tight-binding electrons under a staggered-modulated
magnetic field has been realized~\cite{Iye}, and the Little-Parks
oscillations exhibit the edge part of the corresponding Hofstadter
spectrum.

Motivated by the above experimental and theoretical studies of the
IQHE and the SFP, we investigate the magneto-transport properties
of tight-binding electrons on a 2D square lattice in the presence
of a staggered-modulated magnetic field which has the alternate
flux strength $\phi\pm\Delta\phi$ in neighboring plaquettes.
Hofstadter's butterflies of similar systems have been studied
numerically~\cite{Shi,Oh}. Here, we find a new peculiar
$\Delta\phi$-dependent redistribution behavior of Chern numbers of
Landau subbands and propose that this effect be an experimental
probe to reveal the possible internal orbital-antiferromagnetic
structure of the SFP order in some
systems~\cite{Chakravarty,Chandra}.

\begin{figure}[!htb]
  \vspace{-2.3in}
  \hspace{0.2in}
  \includegraphics[scale=0.35]{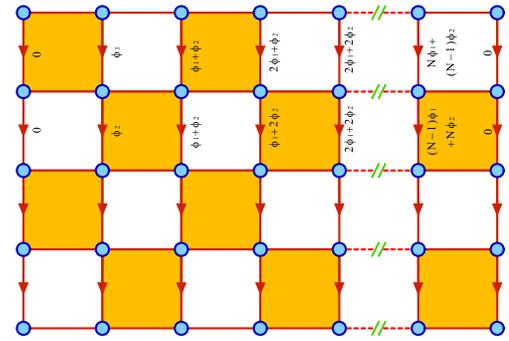}
  \vspace{-0.15in}
  \caption{(color online). Illustration of the selected gauge. $\phi_1=\phi+\Delta\phi$ and $\phi_2=\phi-\Delta\phi$} \label{f.1}
\end{figure}

{\it Formulation.---}The tight-binding model Hamiltonian is given
as follows:
\begin{equation}
H=-\sum_{\langle{ij}\rangle}e^{ia_{ij}}c^{\dagger}_{i}c_{j}+\text{H.c.}+\sum_{i}w_{i}c^{\dagger}_{i}c_{i}
\label{e.1}
\end{equation}
where the hopping integral $t$ is taken as the unit of energy,
$c_i$ ($c^{\dagger}_i$) is an electron annihilation (creation)
operator on site $i$, and $\langle{ij}\rangle$ refers to two
nearest neighboring sites. The magnetic flux per plaquette (the
summation of $a_{ij}$ along four links around a plaquette) is
given by $\phi_1=\phi+\Delta\phi$ and $\phi_2=\phi-\Delta\phi$
alternatively in neighboring plaquettes.
$\phi={p\over{2N}}\times{2\pi}$ ($p$ and $2N$ are coprime
integers) measures the uniform-flux strength, $\Delta\phi$
measures the staggered-flux strength, and both are in units of
$\phi_0/2\pi$ ($\phi_0=hc/e$ is the flux quantum). $w_i$ is a
random potential uniformly distributed between $[-W/2,W/2]$.

The gauge selection for this staggered-modulated magnetic field is
shown in Fig. \ref{f.1} and the corresponding periodical boundary
condition is adopted. In the absence of disorder ($W=0$), the size
of the chosen magnetic unit cell is $2N\times 2$ for
$\phi={p\over{2N}}\times{2\pi}$. Given $N$, the variation of
$\Delta\phi$ does not change the periodicity of the system. In the
presence of disorder ($W\neq0$), the sample considered has the
$2N\times 2N$ unit cell for $\phi={p\over{2N}}\times{2\pi}$ and
the periodical boundary condition is also adopted.

After the numerical diagonalization of the Hamiltonian
[Eq.~(\ref{e.1})], the total Hall conductance for the system can
be calculated through the standard Kubo formula:
\begin{eqnarray}
{\sigma}_{\rm H}(E)&=&{ie^2\over{A_0\hbar}}\sum_{\varepsilon_m<E}\sum_{\varepsilon_n>E}\nonumber\\
& &{{{\langle m|v_x|n\rangle\langle n|v_y|m\rangle -\langle
m|v_y|n\rangle\langle
n|v_x|m\rangle}}\over{{(\varepsilon_m-\varepsilon_n)}^2}}
\label{e.2}
\end{eqnarray}
where $A_0$ is the area of the system, $E$ is the Fermi energy,
$\varepsilon_m$ and $\varepsilon_n$ are the corresponding
eigenvalues of the eigenstates $|m\rangle$ and $|n\rangle$. The
velocity operator is defined as
$v_{\tau}=i\sum_i(c^{\dagger}_{i+\tau}c_ie^{i\phi_{i+\tau,i}}-c^{\dagger}_ic_{i+\tau}e^{-i\phi_{i+\tau,i}})$
with $\tau=\hat{x}$ or $\hat{y}$. And we can rewrite $\sigma_{\rm
H}$ as $\sigma_{\rm H}(E)=e^2/h\sum_{\varepsilon_m<E}C_m$, where
$C_m$ is the Chern number~\cite{Thouless,Arovos} of the $m$-th
subband.

\begin{figure}[!htb]
  \vspace{-0.45in}
  \hspace{-0.25in}
  \includegraphics[scale=0.87]{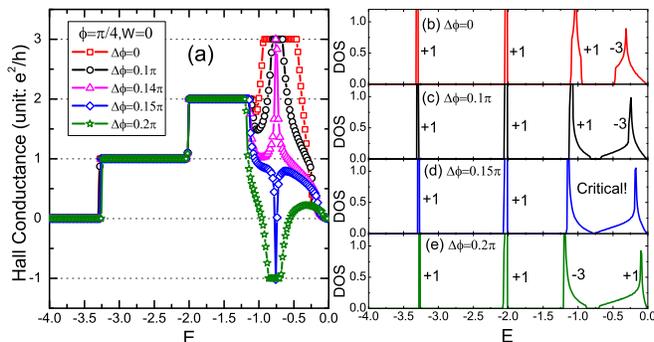}
  \vspace{-0.5in}
  \caption{(color online). The case with $p=1$, $N=4$ and $W=0$.
  (a) The Hall conductance $\sigma_{\rm H}$ versus the Fermi energy $E$ for various $\Delta\phi$'s.
  (b)-(e) The DOS of the cases in (a). The Chern numbers of subbands are shown.} \label{f.2}
\end{figure}

{\it Absence of disorder. (a) An example with $p=1$ and
$N=4$.---}An overall picture for the Hall conductance calculated
by Eq.~(\ref{e.2}) is shown in Fig.~\ref{f.2}(a) with the
uniform-flux strength $\phi={1\over{8}}\times{2\pi}$ and various
staggered-flux strengths
$\Delta\phi=0,0.1\pi,0.14\pi,0.15\pi,0.2\pi$ (only $E\leq0$ part
is shown here). For the uniform case ($\Delta\phi=0$), three
well-defined IQHE plateaus at $\sigma_{\rm H}=l e^2/h$ ($l=1,2,3$)
are clearly shown, corresponding to four Landau subbands [see the
corresponding DOS in Fig.~\ref{f.2}(b)] centered at the jumps of
the Hall conductance at $E\leq0$. With $\Delta\phi$ increasing
from $0$ to $0.2\pi$, one sees a systematic evolution of these
IQHE plateaus. At $\Delta\phi=0.1\pi$, the $l=3$ IQHE plateau
narrows from both sides, and the $l=2$ IQHE plateau narrows from
the right side. At $\Delta\phi=0.2\pi$, the $l=3$ IQHE plateau
disappears and is replaced by a $l=-1$ IQHE plateau.
$\Delta\phi=0.15\pi$ is close to the critical point where the
transition from the $l=3$ IQHE plateau to the $l=-1$ IQHE plateau
takes place. All the values of the Hall conductances in
Fig.~\ref{f.2}(a) fall back to zero when the Fermi energy
approaches $E=0$, due to the particle-hole symmetry.

Such a behavior can be explained through the evolution of the
density of states (DOS) depicted in Fig.~\ref{f.2}(b)-(e). In
Fig.~\ref{f.2}(b),(c) ($\Delta\phi=0,0.1\pi$), there are four
separated subbands (each totally-filled subband contributes
${1\over{8}}$ to the filling factor) at $E\leq0$ with the
corresponding well-defined four sequential Chern integers
$\{+1,+1,+1,-3\}$. In Fig.~\ref{f.2}(d) ($\Delta\phi=0.15\pi$),
the upper two subbands merge together and form a pseudogap which
makes the corresponding quantization of Hall conductance not well
defined. In Fig.~\ref{f.2}(e) ($\Delta\phi=0.2\pi$), the upper two
subbands separate again and the sequential Chern integers are
redistributed as $\{+1,+1,-3,+1\}$. We can conclude that at the
critical filling $\nu={{3}\over{8}}$, there is a direct transition
of the Hall conductance from the $l=3$ IQHE plateau to the $l=-1$
IQHE plateau when $\Delta\phi$ is increased from $0$ up to a
critical value $\Delta\phi_c$.

{\it (b) Redistribution behavior of Chern numbers.---}At $p=1$, we
have numerically confirmed that the Hall conductances at weaker
flux strengths with $N=7$-$36$ all exhibit the similar features as
those in Fig.~\ref{f.2}(a). Thus in general, given
$\phi={1\over{2N}}\times{2\pi}$, the $N$ sequential Chern integers
$\{+1,\dots,+1,+1,-N+1\}$ will be redistributed as
$\{+1,\dots,+1,-N+1,+1\}$ (one can also say that there is  a
transfer of integer $N$ between the last two Chern numbers) when
$\Delta\phi$ is increased from $0$ up to a critical value
$\Delta\phi_c$. Namely, there is a direct transition of the Hall
conductance from the $l=N-1$ IQHE plateau to the $l=-1$ IQHE
plateau) at the critical filling $\nu_c={{N-1}\over{2N}}$. This is
quite different from the global phase diagram proposed by
Kivelson, Lee and Zhang~\cite{Kivelson} for IQHE which predicts
that the only allowed transitions are the nearest-neighbor
plateau-plateau transitions ($l\rightarrow l\pm1$), and is also
different from the disorder-induced higher IQHE plateau to
insulator transition ($l\rightarrow 0$)~\cite{Sheng}.

In cases with $p>1$, based on a Diophantine
equation~\cite{Thouless}, a pattern of subband structure has been
found by Yang and Bhatt~\cite{Yang2}: the subbands away from the
band center form groups with $p$ subbands each, and the total
Chern number of each group is $+1$, while the subbands near the
band center form a special group with $\text{mod}(N,p)$ subbands.

\begin{table}
\caption{\label{t.1} Redistribution behavior of Chern numbers (of
the $E\leq0$ subbands) for $p=3,5,7,9$ and various $N$'s. Only the
last two or three groups are shown for $N\geq13$.}
\begin{tabular}{c|c|c|c}
\hline \hline
$p$ & $N$ &$\Delta\phi$ & Chern numbers of Landau subbands\\
\hline \multirow{4}{*}{$3$} & \multirow{4}{*}{$8$}
& $0$ & $(-5,11,-5)(-5,11,-5)(-5,3)$\\
& & $0.03\pi$ & $(-5,11,-5)(-5,3,3)(-5,3)$\\
& & $0.1\pi$ & $(3,3,-5)(-5,3,3)(-5,3)$\\
& & $0.2\pi$ & $(3,3,-5)(-5,3,3)(3,-5)$\\
\hline
\multirow{2}{*}{$3$} & \multirow{2}{*}{$10$}
& $0$ & $(7,-13,7)(7,-13,7)(7,-13,7)(-3)$\\
& & $0.2\pi$ & $(-3,-3,7)(7,-3,-3)(-3,-3,7)(-3)$\\
\hline
\multirow{2}{*}{$3$} & \multirow{2}{*}{$11$}
& $0$ & $(-7,15,-7)(-7,15,-7)(-7,15,-7)(-7,4)$\\
& & $0.15\pi$ & $(-7,4,4)(4,4,-7)(-7,4,4)(4,-7)$\\
\hline
\multirow{2}{*}{$3$} & \multirow{2}{*}{$13$}
& $0$ & $\dots(9,-17,9)(9,-17,9)(9,-17,9)(-4)$\\
& & $0.1\pi$ & $\dots(-4,-4,9)(9,-4,-4)(-4,-4,9)(-4)$\\
\hline
\multirow{2}{*}{$3$} & \multirow{2}{*}{$16$}
& $0$ & $\dots(11,-21,11)(11,-21,11)(-5)$\\
& & $0.1\pi$ & $\dots(11,-5,-5)(-5,-5,11)(-5)$\\
\hline
\multirow{2}{*}{$3$} & \multirow{2}{*}{$17$}
& $0$ & $\dots(-11,23,-11)(-11,23,-11)(-11,6)$\\
& & $0.08\pi$ & $\dots(6,6,-11)(-11,6,6)(6,-11)$\\
\hline
\multirow{2}{*}{$3$} & \multirow{2}{*}{$32$}
& $0$ & $\dots(-21,43,-21)(-21,43,-21)(-21,11)$\\
& & $0.05\pi$ & $\dots(11,11,-21)(-21,11,11)(11,-21)$\\
\hline
\multirow{2}{*}{$5$} & \multirow{2}{*}{$12$}
& $0$ & $(5,5,-19,5,5)(5,5,-19,5,5)(5,-7)$\\
& & $0.2\pi$ & $(5,5,-7,-7,5)(5,-7,-7,5,5)(-7,5)$\\
\hline
\multirow{2}{*}{$5$} & \multirow{2}{*}{$13$}
& $0$ & $\dots(-5,-5,21,-5,-5)(-5,-5,8)$\\
& & $0.2\pi$ & $\dots(-5,-5,8,8,-5)(-5,8,-5)$\\
\hline
\multirow{2}{*}{$5$} & \multirow{2}{*}{$14$}
& $0$ & $\dots(-11,17,-11,17,-11)(-11,17,-11,3)$\\
& & $0.2\pi$ & $\dots(-11,3,3,3,3)(3,3,-11,3)$\\
\hline
\multirow{2}{*}{$5$} & \multirow{2}{*}{$17$}
& $0$ & $\dots(7,7,-27,7,7)(7,7,-27,7,7)(7,-10)$\\
& & $0.08\pi$ & $\dots(7,7,-10,-10,7)(7,-10,-10,7,7)(-10,7)$\\
\hline
\multirow{2}{*}{$5$} & \multirow{2}{*}{$33$}
& $0$ & $\dots(-13,-13,53,-13,-13)(-13,-13,20)$\\
& & $0.04\pi$ & $\dots(-13,-13,+20,+20,-13)(-13,20,-13)$\\
\hline
\multirow{2}{*}{$7$} & \multirow{2}{*}{$16$}
& $0$ & $\dots(-9,23,-9,-9,-9,23,-9)(-9,7)$\\
& & $0.08\pi$ & $\dots(-9,7,7,-9,-9,7,7)(7,-9)$\\
\hline
\multirow{2}{*}{$7$} & \multirow{2}{*}{$23$}
& $0$ & $\dots(-13,33,-13,-13,-13,33,-13)(-13,10)$\\
& & $0.06\pi$ & $\dots(-13,10,10,-13,-13,10,10)(10,-13)$\\
\hline
\multirow{2}{*}{$7$} & \multirow{2}{*}{$30$}
& $0$ & $\dots(-17,43,-17,-17,-17,43,-17)(-17,13)$\\
& & $0.04\pi$ & $\dots(-17,13,13,-17,-17,13,13)(13,-17)$\\
\hline
\multirow{2}{*}{$7$} & \multirow{2}{*}{$31$}
& $0$ & $\dots(9,9,9,-53,9,9,9)(9,9,-22)$\\
& & $0.04\pi$ & $\dots(9,9,-22,-22,9,9,9)(9,-22,9)$\\
\hline
\multirow{2}{*}{$9$} & \multirow{2}{*}{$20$}
& $0$ & $\dots(9,9,-31,9,9,9,-31,9,9)(9,-11)$\\
& & $0.06\pi$ & $\dots(9,-11,-11,9,9,-11,-11,9,9)(-11,9)$\\
\hline
\multirow{2}{*}{$9$} & \multirow{2}{*}{$22$}
& $0$ & $\dots(5,5,5,5,-39,5,5,5,5)(5,5,5,-17)$\\
& & $0.06\pi$ & $\dots(5,5,5,-17,-17,5,5,5,5)(5,5,-17,5)$\\
\hline
\multirow{2}{*}{$9$} & \multirow{2}{*}{$29$}
& $0$ & $(13,13,-45,13,13,13,-45,13,13)(13,-16)$\\
& & $0.04\pi$ & $(13,-16,-16,13,13,-16,-16,13,13)(-16,13)$\\
\hline
\multirow{2}{*}{$9$} & \multirow{2}{*}{$38$}
& $0$ & $(17,17,-59,17,17,17,-59,17,17)(17,-21)$\\
& & $0.03\pi$ & $(17,-21,-21,17,17,-21,-21,17,17)(-21,17)$\\
\hline \hline
\end{tabular}
\end{table}

At $p=3$ and various $N$'s, through numerical calculations, we
find a more complicated redistribution behavior of Chern numbers
(Table~\ref{t.1}). For instance, we take the case $N=8$, when
$\Delta\phi$ varies from $0$ to $0.2\pi$, the eight sequential
Chern numbers $\{(-5,+11,-5),(-5,+11,-5),(-5,+3)\}$ (each pair of
parentheses represents a group according to the grouping scheme by
Yang and Bhatt~\cite{Yang2}) are redistributed as
$\{(+3,+3,-5),(-5,+3,+3),(+3,-5)\}$. We note that, in each group,
one Chern number is subtracted by $8$ and another neighboring
Chern number is added by $8$. Hence, the total quantized Hall
conductance exhibits a direct transition by $\pm{8}e^2/h$ at the
critical fillings $\nu_c={1\over{16}}$, ${5\over{16}}$ and
${7\over{16}}$. At $N=10$, there is only one Chern number in the
last group which remains unchanged when $\Delta\phi$ varies, while
the other three groups change similarly as the case $N=8$. We have
numerically confirmed that for $p=5$,  $7$ and $9$
(Table~\ref{t.1}), there are also redistribution behaviors similar
to $p=3$.

In general, at $p\geq3$, there is a redistribution behavior of
Chern numbers among each group [except the last group when
$\text{mod}(N,p)=1$ and there is only one Chern number within it]:
a transfer of integer $N$ between two neighboring Chern numbers,
and the redistributed Chern numbers of neighboring groups away
from the band center are antisymmetric in sequence. Hence, the
total quantized Hall conductance will exhibit a direct transition
by $\pm{N}e^2/h$ at critical fillings.

\begin{figure}[!htb]
  \vspace{-0.4in}
  \hspace{-0.2in}
  \includegraphics[scale=0.8]{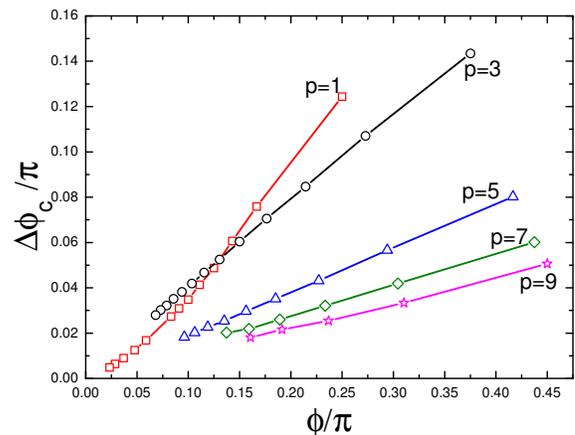}
  \vspace{-0.5in}
  \caption{(color online). $\Delta\phi_c$ versus $\phi$ for $p=1$, $3$, $5$, $7$, $9$ and various
$N$'s (at the critical filling $\nu={{N-1}\over{2N}}$)}
\label{f.3}
\end{figure}

{\it (c) Scaling properties.---}Both the uniform-flux strength
$\phi={p\over{2N}}\times{2\pi}$ at $N=4$-$38$ and the
staggered-flux strength in the previous calculations are
relatively high fields. In order to meet the requirements of
possible experimental realization, we will reduce the flux
strength further. The scaling behavior of $\Delta\phi_c$ with
$\phi$ is shown in Fig.~\ref{f.3}. For $p\geq3$, we concentrate on
the cases where there are two Chern numbers in the last group and
consider the critical filling $\nu={{N-1}\over{2N}}$. For a fixed
$p$, $\Delta\phi_c$ scales with $\phi$ almost linearly. The slopes
of the lines with different $p$'s differ from one another. With
$\phi$ deceasing, all lines approach the origin point, and
therefore $\Delta\phi_c$ approaches zero at the weak-field limit.

{\it Presence of disorder.---}In realistic solids, impurity or
phonon-induced disorder always has effect on the magneto-transport
properties. We now consider $W=1$ to see the influence of weak
disorder on the critical transition behavior.

\begin{figure}[!htb]
  \vspace{-0.35in}
  \hspace{0.0in}
  \includegraphics[scale=0.7]{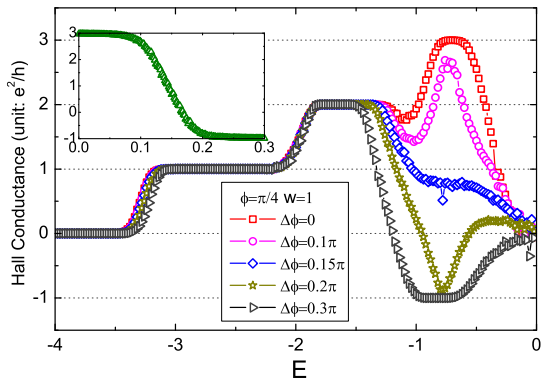}
  \vspace{-0.35in}
  \caption{(color online). The case with $p=1$, $N=4$ and $W=1$
  (1000 random-potential configurations of the lattice size $8\times8$).
  $\sigma_{\rm H}$ versus $E$ for various
$\Delta\phi$'s. The inset shows the evolution of $\sigma_{\rm H}$
versus $\Delta\phi$ at $\nu={3\over 8}$.} \label{f.4}
\end{figure}

As for $p=1$ and $N=4$, we can see from Fig.~\ref{f.4} that the
presence of weak disorder does not smear out the transition of the
Hall conductance from the $l=3$ IQHE plateau to the $l=-1$ IQHE
plateau near the critical filling $\nu={{3}\over{8}}$.

\begin{figure}[!htb]
  \vspace{-0.4in}
  \hspace{-0.0in}
  \includegraphics[scale=0.8]{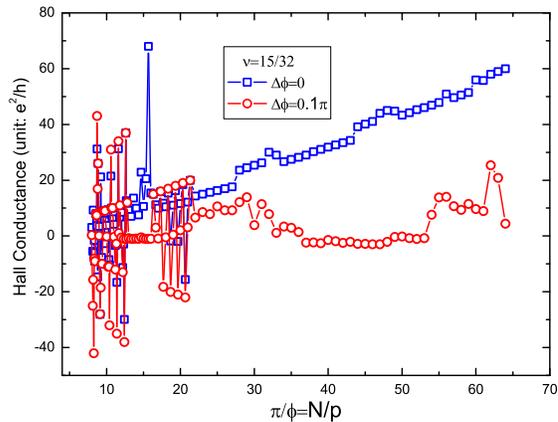}
  \vspace{-0.5in}
  \caption{(color online). At $\nu={{15}\over{32}}$,
  $\sigma_{\rm H}$ versus $\pi/\phi$ for $\Delta\phi=0$ and $0.1\pi$. $N$ ranges
  from $8$ to $64$.
  } \label{f.5}
\end{figure}

{\it An experimental probe of the SFP order.---} As one
application of our theoretical concern, we would take the
staggered-flux part $\pm\Delta\phi$ as the possible internal
orbital-antiferromagnetic structure of the SFP order in some
strongly-correlated electron systems~\cite{Chakravarty,Chandra},
and the uniform-flux part $\phi$ as an external experimental
probe. In Fig.~\ref{f.5}, at the electron filling
$\nu={{15}\over{32}}$, we fix the value of $\Delta\phi$ and do
such a hypothetical experiment by measuring the Hall conductance
which varies with $\phi$. The two curves of the Hall conductances
with $\Delta\phi=0$ and $\Delta\phi=0.1\pi$ exhibit large
deviations at many values of $N/p$. Therefore, such an experiment
could tell us the characteristic differences between the states
with and without the SFP order.

{\it Summary and discussion.---}As for the tight-binding electrons
on a square lattice under a staggered-modulated magnetic field
with the strength $\phi\pm\Delta\phi$, a new peculiar behavior of
the Hall conductances has been found: (a) if
$\phi={p\over{2N}}\times{2\pi}$ ($p$ and $2N$ are coprime
integers) is fixed and $\Delta\phi$ is increased from $0$ up to a
critical value $\Delta\phi_c$, the Chern numbers of Landau
subbands will be redistributed between neighboring pairs; (b) and
hence the total quantized Hall conductance exhibits a direct
transition by $\pm{N}e^2/h$ at one or several critical fillings;
(c) the neighboring two subbands merge together and form a
pseudogap when $\Delta\phi=\Delta\phi_c$; (d) at a fixed $p$,
$\Delta\phi_c$ scales with $\phi$ almost linearly; (e) weak
disorder does not smear out the peculiar transition of the Hall
conductance.

To observe this effect under an accessible uniform magnetic field
with $N\sim1000$, we estimate that both the strength of disorder
$W$ and the temperature $k_{\rm B}T$ should not exceed
${{4t}\over{N}}\sim{{4t}\over{1000}}$ (where $4t$ is half of the
original band width). Taking $t=0.01$ eV, the temperature $T$
should be lower than $0.5$ K.

As for high-$T_{c}$ cuprates, the strength of the internal
staggered magnetic field is estimated to be about $0.1$
T~\cite{Hsu}. One may suspect that: will an internal staggered
magnetic field survive under a stronger external uniform one?  We
believe it does, basing on the numerical results: an external
uniform magnetic field will suppress the $d$-wave superconducting
order, while the SFP order (or equivalently, the DDW order) can
exist or even be enhanced in the vortex cores~\cite{QHWang,JAn}.

This work was supported by the State Key Program for Basic
Researches of China under Grant No. 2006CB921802 and the Research
Grants Council of Hong Kong.

\end{document}